\documentstyle[11pt]{article}

\begin{document}
%

\begin{center}
{\large \bf Cosmology: Neutrinos as the Only Final Dark Matter}
\vskip.5cm

W-Y. Pauchy Hwang\footnote{Email: wyhwang@phys.ntu.edu.tw} \\
{\em Asia Pacific Organization for Cosmology and Particle Astrophysics,\\
Center of Theoretical Sciences, Institute of Astrophysics,\\
 and Department of Physics, National Taiwan University, \\
     Taipei 106, Taiwan}
\vskip.2cm


{\small(November 30, 2010; revised: April 19, 2011;
September 15, 2011; October 20, 2012; August 25, 2013; January 13, 2016)}
\end{center}

\begin{abstract}

Even though neutrinos and antineutrinos are everywhere
in the Universe, their critical importance might be
overlooked, especially because that at least one species
of neutrinos has the mass $0.058\, eV$, far larger than the
cosmic thermalization temperature $1.9^\circ\,K$. The
non-zero mass makes neutrinos participate the galaxy formation 
from the very beginning, in view of the process of
clustering. Unlike the cosmic microwave background (CMB), the
cosmic background neutrinos (CB$\nu$) cannot be uniform. Thus,
we wish to examine the questions such as: Is there some new
source for neutrinos or antineutrinos, that might be detectible
experimentally? Is there some new interaction of neutrinos
with the visible world, that may be of numerical importance
at, e.g., the ultra high energies ($\ge 10^{13}\,eV$)? One
major conclusion is that, on the basis of the Standard Model,
neutrinos would eventually become the {\it only} dark-matter
species left in our Universe.

Our Cosmos is limited in energy for various particles,
electrons or photons without threshold, while protons or
neutrinos having the following hurdles to overcome in reaching
extreme energies such as $10^{18}\, eV$.
In an electron-rich medium, the threshold is $10^{15}\, eV$
for an ultra high energy (UHECR) proton, due to $p + e^-
\to n + \nu_e$. On the other hand, the cosmic background
neutrinos would cut off UHECR neutrinos of greater than
$10^{13}\, eV$ if at least one kind of neutrinos has the
mass $0.05 \, eV$ (as suggested by the experimental value
of $0.058\, eV$), due to $\nu +{\bar \nu}_{CB} \to e^- +
e^+$; this, plus the clustering due to mass, gives us some
hope that this effect might be detectible.

\bigskip

{\parindent=0pt PACS Indices: 98.80.-k (Cosmology); 12.60.-i (Models beyond
the standard model); 12.10.-g (Unified field theories and models)}
\end{abstract}

\section{Prelude No.1}

What is our Universe? When trying to study our Universe
(i.e., Cosmology), we should clarify what the Universe is all
about. We propose that we live in the quantum 4-dimensional
Minkowski space-time \cite{HwangWYP}. The physics in the
entire 20th Century was dominated by both the relativity
principle and the quantum principle, the so-called two
pillars of modern physics. In the beginning of the 21st
Century, we should be ready to admit that the Cosmos that
we are living is the quantum 4-dimensional Minkowski
space-time, as the two pillars of the 20th Century are
telling us.

What do we see in this Universe? Starting from the
(quantum) 4-dimensional Minkowski space-time, the complex
scalar field, if alone, cannot exist, because of the
dimensionless self-repulsive $\lambda (\phi^\dagger \phi)^2$
interaction \cite{Fields}. It is dimensionless, so the
parameter $\lambda$ is {\it not} determined by the
field itself; rather, determined by the (quantum)
4-dimensional Minkowski space-time as a whole.

What do we see in this Universe? Einstein relation,
$E^2={\vec p\,}^2+m^2$, is rather basic (in the
description of the dynamics of motions) and so is its
Dirac linearization, $E={\vec \alpha}\cdot{\vec p} +
\beta m$. Thus, we see electrons, muons, quarks,
etc.; in detail, we see the lepton world, of atom
sizes, and we also see the quark world, of the
$(fermi)^3$ sizes. They are described by the Dirac 
equations of some sort.

Why do we have the faith in this Standard Model
\cite{Hwang417}? Apart from the "ignition" term, it is
completely dimensionless theory - all couplings are
dimensionless, everywhere in the "background" of our
world, the lepton world, and the quark world; they are
determined by the quantum 4-dimensional Minkowski
space-time {\it as a whole}, rather than by its details.
The "ignition" indicates the switch-on of the phase
transition into the "mass" phase (in which particles
have masses while, without the "ignition" term, there
is no mass term in the theory). Or, the "ignition" means
that the God turns on the spontaneous symmetry breaking
(SSB) regarding the mass generation.

Thus, we believe that we live in the (quantum) 4-dimensional
Minkowski space-time with the force-fields gauge-group
structure $SU_c(3) \times SU_L(2) \times U(1) \times
SU_f(3)$ built-in from the very beginning. This is the
"background" of everything. The gauge-group background
structure should be there, {\it a priori}. Otherwise,
conceptually we would not know where it (the force-fields
gauge-group structure) comes from. Experimentally, the
$3^\circ\,K$ cosmic microwave background (CMB) may be
considered as the evidence {\it a priori}. Thus, in this
way we complete the logical description of our Universe.

As for the origin of mass \cite{Origin}, they come from
the spontaneous symmetry breaking (SSB) of the complex
scalar (Higgs) fields; they would disappear altogether
when the temperature is higher than that when the SSB
turns on. All the mass terms would not be there if the
temperature is high enough. Thus, apart from the
"ignition" term, there is no dimensional
coupling (or interaction) in the Standard Model
\cite{Hwang417} - a rather interesting phenomenon
which has some fundamental bearing.

Thus, we already briefly set up the stage of discussing
neutrinos, the species that is the most elusive and that
in fact might control the destiny of our Universe.

\bigskip

\section{Prelude No.2}

Protons and electrons are everywhere in this Universe.
During these days, we are talking about protons of greater
than $10^{18}\, eV$ in the so-called "ultra high energy
cosmic rays" (UHECR's). For instance, AUGER currently is
leading the courageous efforts.

In an anatomy of neutrino oscillations \cite{Hwang6},
we observe that the reaction $p + e^- \to n + \nu_e$
for a proton of energy higher than $10^{18}\, eV$ in
fact will dump the entire energy into the final neutrino,
leaving the final neutron much lower of the energy.
Since the electrons are everywhere in the Universe,
the capture of UHECR protons may decrease the proton
flux in a significant way.

Neutrinos at such UHECR energies might interact much more
easily, like the electrons - the full electroweak
unification making the interactions with neutrinos
the same as those with electrons. This is the reason No.1.
The other reason comes from the exchange of the new
family Higgs bosons - from what we need to understand
neutrino oscillations \cite{HwangYan}, or in the
Standard Model with neutrino oscillations
\cite{Hwang417}.

Nowadays we know pretty well that neutrinos are massive, even
though of some tiny masses. A tiny mass of $0.05\, eV$ would
transform the cosmic background (CB) $1.9^\circ K$ neutrinos
from the relativistic to mass-dominated. We also know very
well that neutrinos oscillate, from one flavor state (which
is {\it not} an eigenstate of the Dirac Hamiltonian) onto the
other, from a linear combination of mass eigenstates onto
the other. For "point-like" Dirac particles doing
oscillations, this interaction causing oscillations must be
understood at the very fundamental level (conceptually).
From the way to write the Dirac equation, there is no size
parameter - thus, from there on, we describe the various
point-like Dirac particles, including leptons and quarks
in the table of "the building blocks of matter".

{\it Moreover, the cosmic background (CB) neutrinos, the cousin
of the cosmic microwave backgrounds (CMB), in fact become
observable. Why? The neutrino mass in the range of $0.05\, eV$
makes the CB temperature $1.9^\circ K$ irrelevant. And the
neutrino mass makes neutrinos to cluster, maybe of the order
$10^5$ or larger. This is a reasonable guess.}

In the language of relativistic quantum mechanics and quantum
fields, if the initial state and the final state are Dirac
particles, what would be in the middle has to be the scalar
field, if we insist on the "renormalizability". The three
entities, the initial and final Dirac fields and the family
Higgs field, each are triplets (in the family space); and
thus they form the cross or curl product. This form is
unique. Consequently, this implies the existence of the
lepton-flavor-violation interaction \cite{Hwang6}.

We know that the minimal Standard Model is a gauge theory based on the gauge
group $SU_c(3) \times SU_L(2) \times U(1)$ with the quark and lepton
multiplets except the right-handed neutrinos. Thus, we may introduce
the other $SU_f(3)$ with the right-handed neutrinos as the triplet - let's
call it the family gauge group. Because so far the family gauge theory is
completely independent of the minimal Standard Model, including the
multiplets used, we could form $SU_c(3)\times SU_L(2) \times U(1) \times
SU_f(3)$, a trivial extension of the minimal Standard Model.

To arrive at an extension\cite{Hwang417} of the minimal Standard Model
that has three generations and that neutrinos oscillate among different
generations, the interaction in the form of Hwang and Yan
\cite{HwangYan} is a necessity. Why does it stop at the lepton
world? The quark world has a much smaller size, of $(fermi)^3$ size,
and, with the $SU_c(3)$ strong interaction, the imposition of
$SU(3)$ constraints might be more than complete.

If we introduce the family Higgs mechanism {\it only} in the lepton
world, it is true that the allowed ranges of the masses of the
family Higgs and of the family gauge bosons are still rather large,
e.g. from a few $GeV$ up to a few $TeV$ at least \cite{Family}. We
note that the conceptual disaster of Landau ghosts associated with
$U(1)$ is no longer there and $SU_f(3)$ makes the lepton world
asymptotically free - the $(123)$ gauge symmetry seems to be a must.

The introduction of the family gauge-field concept in the lepton
world offers an explanation why there are three generations - through
the language of the gauge-field theory. Otherwise, the three
generation, or the famous Rabi's question of muon or lepton
duplication, would remain very mysterious.

To sum up, we choose $(\nu_\tau,\, \nu_\mu,\, \nu_e)$ as
the $SU_f(3)$ triplet, in order to realize the idea. In
completing the story, Hwang and Yan \cite{HwangYan} put the
six left-handed leptons into $\Psi(3,2)$, an $SU_f(3)$
triplet $SU_L(2)$ doublet. The right-handed leptons, i.e.,
the neutrino trio and the charged leptons, both are
$SU_f(3)$ triplets, while singlets under all the other
groups. These are the "basic units" on the "new"
Standard Model \cite{Hwang417}. Because of only the
electroweak data which are relevant, there is so far
not much stringent constraint even from
the precision experiments on the various
couplings for the $SU_f(3)$ gauge symmetry. We name
it as "the lepton world"; our world consists
of the force-fields "background", the quark world, and the lepton
world. The lepton world is of the atomic size, while the quark
world of $(fermi)^3$ size, very different.

In our language, the $SU_f(3)$ gauge sector is the primary dark-matter
world, the 25\% of the present Universe (compared to 5\% for the
ordinary-matter world) - while the body of neutrinos and antineutrinos
is something coupled to it. The $SU_f(3)$ gauge sector is used to
characterize the "family gauge symmetry".

In short, we are proposing \cite{Hwang417} that we live in the
(quantum) 4-dimensional Minkowski space-time with the
force-fields gauge-group structure $SU_c(3) \times SU_L(2)
\times U(1) \times SU_f(3)$ built-in from the very beginning.
Sitting from this "background", we can see the quark world, and
also we can see the lepton world. This serves the platform
for studying neutrinos in cosmology, the main task of this paper.

\bigskip

\section{Neutrino-related Reactions at Extreme Energies}

One should realize that the ultra high energy cosmic rays
(UHECR's) cannot be in the electrons (or positrons), or in
photons if the energy is too high.

The electrons can always be scattered by the
CMB photons, {\it without} a threshold. This
excludes the possibility that the UHECR's be
electrons or positrons.

The high-energy photon can interact with the
cosmic microwave background (CMB) photon,
into the electron-positron pair. That is, for the
reaction $\gamma(k) + \gamma_{CMB}(k') \to e^-(p)
+ e^+(p')$, the energy-momentum conservation
is give by, in the head-on collision,
\begin{eqnarray}
k+ k' &=& E+E'; \nonumber\\
k+(-k') &=& p + p'.
\end{eqnarray}
Since $E=\sqrt{m_e^2+p^2}$, these two equations are
used to solve $p$ and $p'$. The elementary algebra
gives $B^2-4 AC \ge 0$, or the threshold $kk' \ge
m_e^2$.

Using the CMB temperature of $2.7251^\circ \,K$, the threshold
energy is $1.112\times 10^{15}\, eV$. In these days, we are
talking about the UHECR's of $10^{(18-20)}\, eV$. So, the UHECR
photons cannot be there.

So, in our UHECR Universe, we might have protons, stable nuclei,
and (anti-)neutrinos. We assume that stable nuclei, so rare at
these UHECR energies, can be neglected. The left-overs would be
protons and neutrinos.

There is one to generate neutrinos out of the UHECR protons
\cite{Hwang6} - say, the proton of energy of
$10^{18}\,eV$ or higher. One might think that the proton is
stable, i.e., do not decay whatsoever. In the Universe, the
protons would encounter the matter medium and thus the
electrons - then $p(p_1) + e^- (p_2) \to n(p'_1) +
\nu_e(p'_2)$ in the extreme kinematics become possible.
The beam energy-momentum $({\vec P}, E)$ is so huge
compared to $M_W$. The coupling $g^2$, or $e^2$, does
not cut off (the strength) much.

So, we have
\begin{eqnarray}
{\vec P} &&= {\vec P}' + {\vec P}_\nu, \nonumber\\
\sqrt{m^2+P^2} + m_e && = \sqrt{m_n^2 +P'^2}+E_\nu.
\end{eqnarray}
Now, $E_\nu \approx  P_\nu$ because of the tiny neutrino
mass. One obtains
\begin{equation}
P_\nu = N/D, \qquad N=m^2-m_n^2+m_e^2+2m_e P(1+{m^2\over 2P^2}+...),
              \quad D= 2 m_e + ({m^2\over P}+ ...).
\end{equation}

Using the U-gauge (in which the fictitious particles are
absent), the W-propagator is given by
\begin{equation}
D_{\mu\nu}(k) = {1\over i}{\delta_{\mu\nu} + k_\mu k_\nu/m^2_W
\over k^2 +m_W^2 -i \epsilon},
\end{equation}
which is coupled by $j_\mu(k) = <n(p')|J_\mu(0)|p(p)>$
and $j_\nu^{(e\nu)} = i {\bar u}_\nu (k) \gamma_\nu (1+ \gamma_5)
u_e(0)$.

The four-momentum flowing through the weak boson,
${\vec k} = {\vec P}$ and $k_0 = E_\nu + m_e$,
can easily be determined as above.
Thus, we have, for the denominator of the propagator,
$k^2={\vec k}^2-k_0^2\approx - 2 P m_e + ...,$ which
is large compared to $m_W^2$. The numerator is also
controlled by $k_\mu k_\nu/m_W^2$. The numerics on
the numbers larger than $m_W$ is the outcome.

Putting in the masses and the ultra high energy, say,
$P=10^{18}\,eV$, we obtain $P_\nu \approx P$. Thus,
we realize that the flow of the energy-momentum is
through the $W^+$ boson and then, almost completely,
into the neutrino. The final neutron, just like the
initial electron, serves as the spectator, no longer
of initial ultra high energy. Exercise with this
extreme kinematics is lot's of fun!!

This exercise tells us that, at $10^{18}\, eV$,
the transition amplitude is of order unity. The "weak" process
is no longer weak.

In treating this problem, there is certain advantage
in adopting the U-gauge for the leading no-loop
calculations - there is no contribution from the
ghosts. Our results show that our Universe is indeed
rather intriguing, when we consider the UHECR's
behaviors of, e.g., $p+e^-\to n+\nu_e$.

For the ultra high limits such as a proton of
$10^{18}\, eV$ or higher, the process $p+e^- \to
n+ \nu_e$ will help to deliver the proton energy
to the neutrino, in a weak process with the
energy so high that it is no longer "weak".
Of course, this happens in a galactic medium,
where the electron capture would be a norm.
Note that once the energy is dumped into the
neutrino, it disappears, basically. Summarizing
in short, the extreme UHECR kinematics, plus
the lowest-order $W$-graph, suggests that
there may be important mechanisms to cut off
the ultra high energy protons.

In addition, as shown later, such UHECR
neutrinos would be captured by cosmic
background (CB) neutrinos. This effect is
amplified for two reasons: First, the
tiny mass of $0.058\,eV$, much bigger
than the background temperature
$1.9^\circ K$, will take over. Secondly,
there should be clustering of a few orders
due to this mass, different from the situation
when neutrinos are massless.

\bigskip

\section{Dirac Similarity Principle and Minimum Higgs Hypothesis}

One leading question in the physics of neutrinos is whether neutrinos are
Dirac particles, or maybe Majorana particles. A particle that we could
investigate so barely seems to belong to both possibilities. However,
we would argue that neutrinos must be point-like Dirac particles,
satisfying Dirac equation of some sort.

The first argument is as follows: All the building
blocks, such as electrons and quarks, can be described by the Dirac equations
in certain forms and the search of the last forty years for the scalar fields
such as Higgs still remains in vain. By construction, one may get other
relativistic particles but it seems that point-like Dirac particles are
already sufficient. Thus, we try to formulate the "Dirac Similarity Principle"
and the "minimum Higgs hypothesis" as our working hypotheses \cite{Hwang2}
- particularly for the dark-matter world. These are working hypotheses, which
are true so far for eighty years (Dirac) or for forty years and which could
serve as good guidelines when a lot of unknowns (such as dark matter) are at stake.

The second argument is from the mathematics of group theory.
For instance, the left-handed electron and the left-handed neutrino
jointly form a doublet under $SU_L(2)$ - they must have the same
characteristics. If electrons are point-like Dirac particles, but
neutrinos Majorana particles, the underlying group theory is in
error. It's better that neutrinos are also point-like Dirac
particles. In fact, we have been using the group theory in
formulating the Standard Model.

In fact, the Dirac equation comes from the linearization of
Einstein's basic relation $E^2={\vec p\,}^2 + m^2$, a very
elementary demand out of the belief that we live in the
quantum 4-dimensional Minkowski space-time. If some
particle has to be described by something else, the
description may be for a different world than ours
(i.e., the quantum 4-dimensional Minkowski space-time).

Dirac invented the so-called "Dirac equation" to describe the electron, which
later turns out to be the first point-like Dirac particle (in
the Standard Model). For the last eighty years, our searches for point-like Dirac
particles could be summarized by the Standard Model; in fact, in our world, the
point-like Dirac particles belong only to the world of the Standard Model; and in
our space-time they are described as "quantized Dirac fields". On the other hand,
for the last forty years, we were looking for "Higgs particles", spin-zero
quantized Klein-Gordon fields, but surprisingly enough nothing so far. Therefore,
We suggested \cite{Hwang2} that we could formulate our forty-year experience
as the "minimum Higgs hypothesis". In theoretician's search for
the new models, "Dirac similarity principle" and the "minimum Higgs
hypothesis" greatly simplifies the scope. Thinking of 25\% dark
matter, the two empirical working hypotheses should help considerably.
(One is the experience of the last eighty years - what are "point-like Dirac
particles" and are there other point-like configurations that exist in our
space-time? The other is that of the last forty years - why are the
Higgs particles so few?)

Another consideration related to the theory of dark matter is the so-called
"symmetry", including the super-symmetry. In fact, we have a lot of rooms or
loopholes in this regard, but the symmetry considerations should play a major
role in the theory of dark matter. Maybe the "symmetry" is equivalent to
the "interactions" of some form. To this end, we use "Dirac similarity principle"
and the "minimum Higgs hypothesis" as two simplifying working
conjectures.

Of course, we finally think through all these in reaching the final
Standard Model \cite{Hwang417}, where we could understand easily
\cite{Origin} and even understand easily why we live in the quantum
4-dimensional Minkowski space-time \cite{Fields}. The two working
rules did help us to reaching at the status of today.

In retrospect, we should try to say the followings:

Under "Dirac similarity principle" and the "minimum Higgs hypothesis", the extended
Standard Model would be unique if the gauge group is fixed. For example, the extra
$Z^{\prime 0}$ model, for the $SU_L(2)\times U(1) \times U(1)$ gauge group, is unique
\cite{Hwang3}. The group $SU_L(2) \times SU_R(2) \times U(1)$ now gives rise to a
unique left-right symmetric model - one out of the left-right models \cite{Salam},
though we "reject" this model because of its adoption in violation with the spirit
of the "basic units" (in place of "building blocks of matter").

The adoption of "basic units" is based on the viewpoint that
the motion of a particle is described by the sum of the kinetic
energy term and the potential term. There is one, only one,
kinetic energy term for each physical system. The transition
from a particle to a field does not alter this spirit in our
language. This raises an obvious objection against
the left-right model \cite{Salam}.

We have been curious why there are three generations of fermions, i.e.,
quarks, charged leptons, and neutrinos; the so-called "family symmetry" but
without giving a reason. To our knowledge, these particles are all described
by Dirac equation and the so-called point-like Dirac particles (or, quantized
Dirac fields). The similarity to the electron is rather strange and thus I
call it "Dirac similarity principle", although it applicability to the case
of neutrinos is waiting for verification. These might be the only additional
point-like Dirac particles realized in our space-time (as described by
quantized Dirac fields).

We are living in a universe that at this moment there are a lot of
unknowns - about 70\% dark energy, 25\% dark matter, and only 5\% "visible"
ordinary matter. The well-known minimum Standard Model is used
to describe the 5\% "visible" ordinary matter, leaving 95\% of the Universe
untouched. Neutrinos, interacting so weakly with other ordinary matter, in
some sense could be regarded as one kind of dark matter and in this paper
be treated as a messenger between the ordinary matter and the dark matter.

The story may be such that the Standard Model, which describes
an immense amount of observing data, has been substantiated to high precision
and it would be extended somehow to describe the bulk of dark matter. On the
other hand, we all know that the 70\% dark energy might be represented, to the
first approximation, as due to the presence of the cosmological
constant. Accordingly, we are left with the dark matter and the ordinary matter,
the so-called "matter", to worry about.

There are some correlations between the dark-matter world and the "visible"
ordinary-matter world - for instance, the Milky Way has about four or five
times in mass of dark matter associated with it, judged from the rotation
curve of the spiral. If there is no interaction, except the gravitation force,
between the dark-matter particles and the ordinary-matter particles, then
such correlations should not exist. On the other hand, if strong and
electromagnetic interactions exist between the dark-matter particles and the
ordinary-matter particles, it doesn't fit the description of the dark-matter
galaxies or world. So, at best, it seems that the weak interactions could
exist between the dark-matter particle and the ordinary-matter particle.

The fact that the electromagnetic and weak interactions are unified into
the $SU_L(2)\times U(1)$ theory puts an important constraint - the dark-matter
particles have to be singlets under $SU(2) \times U(1)$ or to be at most
$(I_W,Y)=(0,0)$ members (neutrinos). So, that dark-matter particles don't
participate in strong and electromagnetic interactions put a severe constraint
on the identity of the dark-matter particles - the only left-over in the
ordinary-matter world would be neutrinos and antineutrinos. That there is
some galactic correlation as mentioned above indicates that neutrinos are also
one kind of dark-matter particles - otherwise, there would be no communication
with the dark-matter world {\it at all}.

{\it To sum up on the 25\% dark matter, neutrinos (antineutrinos)
are the only particles which can interact with other dard-matter
particles, even though very feeble. Our arguments all point to
this conclusion.}

Could neutrinos be the messenger between the ordinary matter and the dark matter?
Could neutrinos (in the ordinary-matter world) interact with the species in the
dark-matter world? In fact, the neutrinos do not fit squarely into the
minimum Standard Model - the Model says that they should be massless but the
experiments tell not. In other words, the minimal Standard Model needs to be
extended somehow. That is why we have proposed to extend the Standard
Model \cite{Hwang417}, that would make the neutrino sector much more
interesting.

In other words, we think that "neutrinos are also one kind of dark matter" -
neutrinos also interact with other dark-matter particles or neutrinos also have
connections with dark-matter interactions. By this assertion, we
rule out the dark-matter candidacy of the charge leptons, such as electrons, and
of the quarks, thinking of these ordinary-matter particles that would be too
visible. We suspect that there is indeed some bridge, such as neutrinos, between
dark matter and ordinary matter, since it is believed that dark matter is
clusterized near the visible world.

There is a scenario for clustering - for ordinary matter, we know that they are
clusterized into galaxies, clouds, etc. while for dark matter they might be
clusterized into invisible galaxies, clouds, etc., maybe of order ten or larger
(in length). Neutrinos have tiny mass in the sub-eV range or the feeble
interactions effectively of the sub-eV range - it fits the description.
This in fact may explain why our Milk Way or the other galaxies has a large
spiral arm. So, the invisible dark-matter "galaxies", of size $10^3$ or bigger,
serve as the hosts of the ordinary galaxies, such as the Milk Way.

\bigskip

\section{Cosmos: Neutrinos in the Peculiar Bridging Role}

Consider the Universe consisting of protons, electrons,
photons, neutrinos, and their antiparticles. Consider the
ultra high energy cosmic rays (UHECR's) limits of the
system. Protons of the UHECR energies would be killed by
the capture reaction $p+ e^- \to n + \nu_e$ at
$10^{15}\,eV$. Electrons would be killed by the Compton
scattering $e^-+ \gamma_{CMB} \to e^- + \gamma$ or by
the pair-annihilation process $e^-+ e^+ \to \gamma +
\gamma$, or $e^- + e^+ \to \nu +{\bar \nu}$, all
without the threshold. In fact, the cosmic background (CB)
neutrinos also provide another hurdle for electrons
and protons - also without the threshold. If
one of the neutrino species has a tiny mass such as
$0.05 \, eV$ (a realistic value), then the clustering
effect would be orders bigger (such as $10^5$) and
the mass hindrance effect from cosmic background
neutrinos would also be replacing the original
$1.9^\circ K$ temperature effect.

So, this Universe would eventually tend to be very
smooth in the UHECR's energy limit; that is, virtually
no particles of extremely UHECR's energies.

Thus, the neutrino-lization of all very high energy protons,
of energies greater than $10^{18}\, eV$, helps the Cosmos
to smoothen out the content at the UHECR energies. Most UHECR
energies are with neutrinos, which are in turn under
${\bar \nu} + \nu_{CB} \to e^- + e^+$ with a threshold of
$10^{13}\, eV$ (provided $m_\nu= 0.05\, eV$), see the next
section. There are no reactions in strong interaction and in
electromagnetic interaction, the quiet Universe!!

Returning to the Standard Model \cite{Hwang417}, the lepton world
is described by the lagrangian:
\begin{eqnarray}
{\cal L} = & - {\bar R} \gamma_\mu \{ \partial_\mu - i \kappa
{\lambda^a \over 2} F^a_\mu + i g' B_\mu \} R \nonumber\\
& - {\bar L} \gamma_\mu \{ \partial_\mu - i \kappa {\lambda^a\over 2}
F^a_\mu - i g {\tau^i\over 2} A^i_\mu + i {g'\over 2} B_\mu  \} L\nonumber\\
& - {\bar N} \gamma_\mu \{ \partial_\mu - i \kappa {\lambda^a\over 2}
F^a_\mu \} N,
\end{eqnarray}
with $L$ the left-handed lepton $SU_f(3)$ triplet and
$SU_L(2)$ doublet, $R$ for the right-handed $SU_f(3)$
triplet charged leptons and $N$ for the right-handed
neutrinos. This lagrangian is basically the "old" Standard
Model plus the $SU_f(3)$ gauge theory. For notations,
please refer to \cite{Books}.

Using the above lagrangian, we write down the couplings of
the dark-matter particles.

(1) Neutrino-familon interactions:
\begin{equation}
i \kappa \bar\Psi \gamma_\mu {\lambda^a\over 2} F_{\mu}^a(x) \Psi,
\end{equation}
with $\Psi$ standing for $R$, $N$, or $L$.
Here $F_\mu^a$ are the eight family gauge bosons (i.e., familons).

In addition, there are couplings directly to the family Higgs:

(2) Neutrino-Higgs (mass) interactions \cite{HwangYan}:
\begin{equation}
  i {h\over 2} \bar\Psi_L(3,2) \times \Phi(3,2)
  \cdot \Psi_R(3,1) + h.c.,
\end{equation}
where $\Phi(3,2)$ is the triplet Higgs field which makes all
familons massive. On the notations, $\Psi_L(3,2)$ is $L$,
$\Psi_R(3,1)$ is $N$, and so on. $h$ is very small, explaining
why neutrinos have tiny masses.

(3) The feeble reactions brought in by the charged Higgs:
\begin{equation}
 i {h^C\over 2} \bar \Psi_L(3,2) \times \tilde \Phi(3,2)
 \cdot \Psi_R^C(3,1) + h.c.,
\end{equation}
where $\Psi^C(3,1)$ consists of the charged leptons.
It is in the form similar to the neutrino-Higgs mass interactions,
leading to the decays such as $\phi^\pm \to \mu^\pm +
\nu ({\bar \nu})$. $h^C$ are much larger than $h$ since
it explains the splitting in the masses of charged leptons.

The above three couplings are all those which we have - they are
renormalizable, and they are dimensionless. There is nothing more;
so, they decide how the dark-matter particles behave (and evolve)
in the Standard Model \cite{Hwang417}.
These couplings are dimensionless, so that they are determined by
the 4-dimensional Minkowski space-time, {\it not} by the fields
themselves \cite{Hwang417}.

The most peculiar part of the story is as follows: These dark-matter
reactions are extremely narrow, if they could have been
seen. But they cannot be seen except the charged channels such as
$\phi^+ \to \mu^+ + \nu$. They are extremely narrow due to lack of
three-body, or more-body, channels. Thus, this world is really
"dark". In our Cosmos, these reactions are very important but they
cannot be seen except the charged channels that might be
barely studied.

Neutrino-familon gauge-coupling interactions and neutrino-Higgs
mass interactions involve two-body decays; they can be treated
easily on the theoretical side but most of them cannot be observed
on the experimental ground; most of them, like neutrinos, are neutral
and participate only weak and feeble family interactions. They
may be studied for the higher-order effects; or, studied indirectly
through the feeble weak interactions brought in by charged Higgs
$\phi^\pm(3,2)$. [Item (3) above.] Thus, most of the dark channels
are completely dark.

We are confident that the Standard Model \cite{Hwang417} is unique.
The three complex scalar fields $\Phi(1,2)$ (SM-Higgs), $\Phi(3,1)$
(purely family Higgs), and $\Phi(3,2)$ (mixed family Higgs) are there
because they are "related" and their mutual interactions are enough
to overcome each of the self-repulsive $\lambda$ interactions
\cite{Fields}. The Dirac's linearization of the Einstein basic
relation, $E^2 = {\vec p\,}^2 + m^2$, means that there are only a few
Dirac fields (point-like Dirac particles). The force-fields
gauge-field structure $SU_c(3) \times SU_L(2) \times U(1) \times
SU_f(3)$ is probably "built-in from the outset" with the
Lorentz-group structure - the (quantum) 4-dimensional Minkowski
space-time, the place in which we are living. The uniqueness of
the Standard Model \cite{Hwang417} implies that eventually the
neutrinos will be the only final dark-matter particles.

{\it From the most peculiar part of the story for the Standard
Model \cite{Hwang417}, almost all decays are invisible. The
direct experimental tests must come from decays of the charged
Higgs $\Phi^+(3,2)$, such as $\phi^+\to \tau^+ + \nu$, etc., sudden
appearance of a charged heavy lepton and nothing else. The charged
family Higgs $\Phi^+(3,2)$ is only a peripheral byproduct of the whole
idea. Of course, there are numerous indirect experimental tests of the
invisible parts of the Standard Model \cite{Hwang417}.}

Of course, for the three dark-matter couplings, we do not know
that whether they are all real or some of them are complex,
remembering that CP violations do allow the presence of
imaginary numbers. The above dark-matter interactions,
presumably fairly weak, occur in the neutrino sector
and differentiate neutrinos - they are invisible and
they can be tested {\it indirectly} via experiments.
The $SU_f(3)$ gauge coupling
$\kappa$ is anticipated to be considerably smaller than the
electroweak coupling $g$ ($=0.6300$) - say, $\kappa=0.1$
in our example. On the other hand, $h\cdot u_i$ determines the
tiny neutrino masses - which are in sub-eV's or smaller.
(That helps to give an estimate, $h\sim (10^{-2} - 10^{-4})$.)
Also, we note that $h^C$ for charged leptons is quite normal,
yielding the normal mass splitting for charged leptons.

If the family gauge group is $SU(3)$ as in \cite{Hwang417}, the eight gauge
bosons $F_\mu^a$ and the pair of the family Higgs triplets $\Phi(3,2)$ and
$\Phi(3,1)$, together with the three neutrinos (and anti-neutrinos), would
serve the main body of the dark-matter particles. Except the known neutrinos,
these unknown family particles are introduced to be reasonably massive,
presumably about $\sim (5 - 100)\,GeV$ (in the lepton world) - but they cannot
be accessed by the Large Hadron Collider (LHC). They may decay, relatively
rapidly, through the invisible modes, except $\phi^+ \to \tau^+ +\nu$, etc.
as mentioned before.

Intuitively, years ago we began to "formulate" our working
conjecture about the
"minimum Higgs hypothesis" as follows: There should be the minimum
number of Higgs multiplets and the couplings to the "remote" Higgs
should be much "smaller" compared to the leading Higgs (such as the
Standard Higgs doublet). However, the gauge-group structure $SU_c(3)
\times SU_L(2) \times U(1) \times SU_f(3)$ seems to be born with
the 4-dimensional Minkowski space-time - so, the group index is
natural with the identity of an object. It is easy to argue that
the Lorentz group exists, that the gauge-group structure
$SU_c(3) \times SU_L(2) \times U(1)$ exists, and that
the gauge-group structure $SU_f(3)$ might also exist. If
something is a singlet under all these, then there are no
interactions between this "something" with our world - this
"something" simply does not exist for us.

Now, in our Standard Model \cite{Hwang417}, we have three complex scalar
Higgs fields, $\Phi(1,2)$ (Standard-Model Higgs), $\Phi(3,2)$
(mixed family Higgs), and $\Phi(3,1)$ (purely family Higgs). The
born self-repulsive interaction $\lambda (\phi^\dagger \phi)^2$
prevents all other "un-related" complex scalar fields from
existence. This gives some explanation on whether the complex
scalar fields exist or not. We have escaped this "minimum Higgs
hypothesis" by realizing that in our world there exist only
three Higgs fields and nothing more.

All the dark familons (i.e., family gauge bosons) or dark Higgs
should not be massless or too light. For example, the cross
sections for neutrino scatterings (off quarks or off charged
leptons) would be modified by their presence (significantly if
some of these dark species are massless or small), so as to
deviate from the known minimal Standard Model results
\cite{PDG}. In other words, any of these masses could be
taken to be $\ge\, 5\, GeV$, on the safe side \cite{Origin}.
Of course, it would be much more fun if these dark matter
particles would be indirectly observed some day.

One effect is the familon-loop corrections in the scattering
of neutrinos off the electron or off the quarks - similar to the so-called
$\rho-$parameter calculation in the minimal Standard Model \cite{PDG}.
Although the $\rho-$parameters for the $\tau-$ and $\mu-$ sectors are
completely unknown - and maybe remain to be completely unknown for a long
time to come, the breakdown of the $\tau-\mu-e$ universality remains
a genuine possibility. On the other hand, the $\rho-$parameter in the
electron case, in view of its relative larger
error, does permit for a larger domain for the familon masses.

The most famous high-order loop effect has to do with $g-2$
\cite{Kinoshita}, for electrons or for muons. Without the detailed
calculations involving the family particles (i.e., familons
and family Higgs), we can't say too much. Since the corrections
from the family gauge bosons and from the family Higgs
would be figured out {\it order by order}, we should anxiously
waiting for the lowest-order family-loop calculations.

In this note, we set out to examine questions associated with Cosmology.
Question No.1 has to do with whether a phase transition might occur
at $10^{-11}$ or $10^{-12}$ sec, or slightly earlier, after the
Big Bang. After all, from our experience with QCD ($SU_c(3)$),
we anticipate that there are rooms for the phase transitions. If so,
what is the major effect of the familon phase
transition(s)? Question No.2 has to do with the fact that all familons
and all family Higgs should eventually decay into lighter species (likely
in neutrinos) - is this some observable effect with the cosmological
time as a measuring stick? These questions are difficult to answer
quantitatively in the Standard Model \cite{Hwang417} as of now
since our understanding of dark matter is still lacking.

We are particularly interested in the aspects with phase
transitions, because, in the origin of mass \cite{Origin},
the masses could be turned off {\it completely}. In the
symmetry-restored phase, every object is without the mass,
or, massless, while after spontaneous symmetry breaking
(SSB) when the "ignition" term is switched on, all the
mass terms suddenly appear. That is, the "ignition" term
plays the role of switching on or off the desired "phase
transition".

\bigskip

\section{From neutrino oscillations to another revolution?}

In an anatomy of neutrino oscillations \cite{Hwang6}, we realize
that the mass eigenstates and eigenvalues can be defined, as by
the solutions to the Dirac equation, but the flavor "eigenstates"
cannot be defined in a similar way - that is, a flavor state can
only be represented as a linear combination of the mass eigenstates.
In other words, the symbol "$\nu_\mu \to \nu_e$" ($\nu_\mu$
oscillating into $\nu_e$), when written in Dirac spinors (mass
eigenstates), in fact does not mean much.

Experimentally, we cannot prepare a beam that is, e.g.,
the muon-like neutrino beam, since it would oscillate away.

This raises a lot of fundamental questions since, we
do calculations on a certain reaction, we could label each
participating particle by its energy and momentum - so,
to respect the energy-momentum conservation. Thus,
neutrino oscillations make this very difficult, if not
impossible, even though the masses are tiny.

{\it Looking at the neutron beta decay $n\to p + e^- +
{\bar \nu}_e$ for example, the energy and momentum of the
electron are fixed if we use Dirac equation for the electron,
so is the same for the neutron, and so is for the proton;
they are fixed for all three of them, how is for the last,
the neutrino since the energy-momentum conservation already
fix up for the neutrino? The law of energy-momentum
conservation is wrong, or the description of the neutrino
via Dirac equation is in trouble, or $U_{e i} v^i(p,s)$
with the Dirac spinor $v^i(p, s)$, or ${\bar \nu}_e$,
a linear combination could act like an entity?}

There is the term "off-the-mass-shell" but which never gets
defined precisely. It would be too big a thing to violate
the law of energy-momentum conservation, but can a properly
defined "off-the-mass-shell" save the conceptual difficulty?

{\it Our resolution for the neutron beta decay would be that
the Dirac linear combination $U_{e i}v^i (p,s)$ can act like
an independent entity.}

Neutrinos enter weak interactions via the flavor states while
the mass terms through the mass eigen-states. The known heaviest
neutrino mass \cite{PDG} is about $0.04\, eV < m_i < (0.2 - 0.4)\, eV$.
This is rather interesting for many reasons. For example, in
the reaction $\nu + {\bar\nu}_{CB} \to e^- + e^+$ by high
energy $\nu$, the CB neutrinos now have the rest mass much bigger
than the temperature equivalent ($1.9^\circ K$), maybe making the
detection of cosmic background neutrinos somewhat easier.

This last example brings in another story - there is $1.9^\circ$
cosmic background neutrinos and they would prevent the UHECR
neutrinos. For the reaction $\nu(k) + {\bar \nu}_{CB}(k') \to
e^- (p_1) + e^+(p_2)$, we have
\begin{eqnarray}
k + \sqrt{m_\nu^2 + k^{\prime 2}} &=& E_1+E_2,\nonumber\\
{\vec k} + {\vec k}' &=& {\vec p}_1 + {\vec p}_2.
\end{eqnarray}
Or, we find
\begin{equation}
k \sqrt{m_\nu^2+k^{\prime 2}} \ge 2 m_e^2.
\end{equation}
Using $m_\nu \sim 0.05 \, eV$, the threshold is about $10^{13}\, eV$,
a fairly unexpectedly-small UHECR threshold.

{\it In fact, this points to a potentially interesting research
problem. Currently, IceCube Collaboration observed two $PeV$
($10^{15}\, eV$) neutrino events \cite{IceCube}. We just said
that the cutoff energy is $10^{13}\,eV$ for the neutrino mass of
$0.05\,eV$. And at least one kind of neutrinos have the mass
$0.058\, eV$. The situation is very interesting, indeed.}

One very important issue is that if neutrinos are massive then these
neutrinos would tend to clusterize, maybe around the galaxies. If
clusterized around the galaxies, the CB neutrino distributions
would deviate from the uniform distributions like in the CMB photons.
As an estimate, we may assume that the neutrinos may clusterize around
the galaxy but not too far, i.e., farther than the nearby galaxies.
So, in a rich galaxies region such as our Virgo cluster, the galaxy
region should occupy only up to one part in $10^5$ (just an estimate).
So, the neutrino density near the galaxy would be enhanced compared to
the average density by a factor of $10^5$ or higher.

The roles played by neutrinos with the tiny masses observed are
of critical importance in our Universe, so much different from the
situations with the massless neutrinos as assumed in old days.
Remember that at least one of the neutrinos has in mass $0.058\,eV$.

For the $\Delta L =0$ oscillations, the formulae read \cite{Kayser}
\begin{eqnarray}
&P(\nu_\alpha\to \nu_\beta)=\delta_{\alpha\beta}-4\sum_{i>j} Re\{U_{\alpha i}^*U_{\beta i}
U_{\beta j}U_{\beta j}^*\}sin^2[1.27\Delta m_{ij}^2(L/E)] \nonumber \\
&\qquad\qquad +2 \sum_{i>j} Im\{ U_{\alpha i}^*
U_{\beta i}U_{\alpha j}U_{\beta j}^*\} sin [2.54 \Delta m_{ij}^2 (L/E)].
\end{eqnarray}
Here $\alpha$, $\beta$ flavor indices, $i$, $j$ the mass eigenstates,
$\Delta m_{ij}^2 \equiv m_i^2 - m_j^2$ is in $eV^2$, $L$ is in $km$,
and $E$ is in $GeV$.

This fix up the experimental set-up for the detection for neutrinos.
For the $^8B$ solar neutrinos as example, we have $E_\nu^{max}=14.06\,MeV$ and
$\Delta m_{i,j}=10^{-5}\to 10^{-3} eV^2$ so that the distance $L$ would be
$(10 \to 1,000) km$ to guarantee the argument of sine or cosine would be
order unity. The other golden number is the cross section of the order
$10^{-42} cm^2$, which is rather small. Thus, it is rather difficult
to detect neutrino oscillations for $MeV$ neutrinos, in fact, for all
neutrinos.

Thus, the above formulae, if neutrinos oscillate into themselves, 
could help us to design certain experiments or to analyze the 
feasibility of a given experiment or of a certain reaction.

Now supposing that the family gauge symmetry might be there
\cite{Family}, as in the Standard Model \cite{Hwang417}, neutrino
oscillations present some conceptual difficulties, as explained
earlier \cite{Hwang6} - such that the flavor states are {\it not} 
eigen-states. The resolution(s) might touch upon the foundation 
of the quantum mechanics, and more. Imagine the dark world with 
only dark-matter neutrinos are floating around; our Universe 
might in fact approach to it some day long time from now.

To close, let us look at the early Universe at time $t=10^{-12} sec$
and at the times earlier. Einstein's General-Relativity and the
simple Equation of States (EOS) for the ordinary matter (using
perfect fluid for example) yield
\begin{equation}
T\approx 1\, TeV, \qquad \rho_\gamma \approx 6.4\times 10^{24} gm/cm^3, \qquad
\rho_m\approx 10^{14} gm/cm^3.
\end{equation}
Analogously, we have, at $t=10^{-13}sec$,
\begin{equation}
T\approx 3.2\, TeV, \qquad \rho_\gamma \approx 6.4 \times 10^{26} gm/cm^3, \qquad
\rho_m \approx 3.2 \times 10^{15} gm/cm^3.
\end{equation}

The condensate squared $w^2$ of the "ignition" channel $\Phi(3,1)$
determines the temperature $T_c^f$ for the familon phase transition.
In our example \cite{Hwang417} with $cos\theta_P=0.6$, we have
$w^2=v^2/6$ or $w\sim 100\, GeV$. So, both the above two temperatures
are earlier than the familon phase transition in the history
of the early Universe.

The atomic scale is about $1.6726\times 10^{-24}gm/(0.529\times 10^{-8}cm)^3$
(proton mass/(Bohr radius)**3) or $11.30\, gm/cm^3$. The atomic scale is where
quantum physics is so prominent that we first learnt quantum mechanics from
there. The above numbers are comparable to the nuclear matter density
$2\times 10^{14} gm/cm^3$.

Somewhere at some temperature $T_c$, we have the $SU_f(3)$ super phase
transition that generates masses of all particles, except photons
\cite{Origin}. Before (i.e., above $T_c$) this super phase transition,
we would call the vacuum as the "familon vacuum"; after, the "mass
vacuum" sine all particles have masses (except the photons). For
further discussions on phase transitions, one might consult Hwang
and Kim \cite{phase}.

People know a little nuclear physics would start worry about those densities:
A star of a solar mass would collapse into a black hole at about
$\rho_m\approx 10^{16}gm/cm^3$ but with much lower temperatures. How about
the photon sphere with much higher densities, that is greater than
$\rho_\gamma \approx 10^{24} gm/cm^3$? Questions of this kind may in fact
lead to some big revolution eventually.

If the Standard Model \cite{Hwang417} could say anything, the
above temperature, at $t=10^{-13}\,sec$ or at $t=10^{-12}\, sec$,
means that the spontaneous symmetry breaking (SSB) is yet to
take place - and at that time there are {\it no} mass parameters
if you look closely at the Standard Model. So, the mass density,
$\rho_m$, is the wrong way to describe the system. It is clear
that, before SSB, we should, at first, try to clarify what would
be the proper language to use.

In other words, the term "mass" already ceases to have its
meaning when the temperature is rather high and the symmetry
is there (i.e., no longer spontaneously broken) \cite{Origin}.
At such temperatures, the spontaneous symmetry breaking (SSB),
in generating the masses, simply do not turn on \cite{Origin}.
When the matter density or the photon density becomes too high,
certain commonly-used terms lose their meanings and they are
{\it no longer} used. Or, we should
examine the adoption of certain terms under these extreme
situations. In addition, the term "mass" disappears when
the Universe undergoes the SSB phase transition and everything
becomes massless \cite{Origin}. We wish to call this SSB as
the "super phase transition" for the mass generation.

\bigskip

\section{Side Conclusion: There is no particle with excessive energy.}

In our Cosmos, there are a few "stable" entities (particles or
fields), namely, photons, neutrinos, electrons, protons, and
their antiparticles, that their participation decides the
long-term evolution of the entire system.

For photons, the cosmic microwave background (CMB) will cut off
the ultra high energy comic ray (UHECR) photons at the energy of
$10^{15}\,eV$.

For neutrinos, the cosmic background (CB) neutrinos would
cut off the UHECR neutrinos of the energy of $10^{13}\,eV$
if the neutrinos have the mass $0.05\,eV$. Here the mass effect
will take over instead of the CB temperature $1.9^\circ K$
effect. There is also the clustering effect due to the neutrino
mass. Thus, the CB neutrinos may become detectible.

For protons, the electron capture of the UHECR proton would dump
all the energy onto the neutrino, via $p+e^- \to n+ \nu_e$ in the
galactic medium environment.

For electrons or positions, they could capture the CMB photons or
capture the CB neutrinos, in lowering the UHECR energies, without
a threshold.

Our Cosmos is rather peaceful in preventing some particle
from grabbing too much energy for itself.

\bigskip

\section{Conclusion: Neutrinos are the only final dark matter.}

To sum up, in our Universe we see only three complex scalar fields:
$\Phi(1,2)$ (SM Higgs), $\Phi(3,2)$ (mixed family Higgs), and
$\Phi(3,1)$ (purely family Higgs). There is no more complex scalar
field because of the self-repulsive $\lambda (\phi^\dagger \phi)^2$
interaction \cite{Fields}. The three complex scalar fields
are responsible for generating the masses of the various gauge
bosons, of the various Higgs bosons, of the various quarks, and
of the various leptons, upon the spontaneous symmetry breaking
(SSB) \cite{Origin}. In our Universe, we see the quark world, thus
its $(123)$ gauge symmetry, in the $(fermi)^3$ sizes; also, we
see the lepton world, of another $(123)$ gauge symmetry, in the
atomic sizes. Every coupling, except the "ignition" term, is
dimensionless, thus determined by the quantum 4-dimensional
Minkowski space-time {\it as a whole}, rather than by its
detailed content \cite{Hwang417}.

The family Higgs $\Phi(3,1)$ and $\Phi^0(3,2)$ decay dominantly into
"invisible" modes containing neutrinos and antineutrinos. Even if
we might see them in higher orders, it is an impossible task
experimentally. We can try to detect the charged sector $\Phi^+(3,2)$
to verify the existence of the whole idea.

{\it The uniqueness of everything guarantees that eventually neutrinos
would become the only final dark-matter particles. Other dark-matter
particles, including the family gauge bosons and family Higgs bosons,
all will decay away mostly into neutrinos.}

In our Universe, using the Standard Model \cite{Hwang417}, the
familons (family gauge bosons) and family Higgs, i.e., those 
dark particles, all decay away into neutrinos and antineutrinos, 
another dark particles. On the UHECR front, the proton in the 
stellar medium, a fairly visible particle, can transfer
almost the entire energy to the neutrino, via the electron 
capture $p + e^-\to n + \nu$. But cosmic background neutrinos with the
tiny mass of $0.058\,eV$ would prevent neutrino of much too high 
energy (e.g., $10^{13}\, eV$). The dark-matter part of the Universe
is neutrino-lized and thus clustered. The extent of the neutrino
cluster-izations would determine whether the effect could be 
detected. The fate for the visible part of our Universe would be 
in protons, electrons, and (CMB) photons. All of them are cut-off 
at too high UHECR energies.

{\it Thus, we find that, as in the Standard Model
\cite{Hwang417}, all the dark particles decay away except the
neutrinos and antineutrinos, those which are the only final 
dark-matter particles. In general, there are cut-off in the 
UHECR energies, because of the cosmic background neutrinos and
the cosmic microwave background.}

\bigskip

\section*{Acknowledgments}
The Taiwan CosPA project is funded by the Ministry of Education (89-N-FA01-1-0
up to 89-N-FA01-1-5). This research was also supported in part by National
Science Council (NSC99-2112-M-002-009-MY3).

\end{document}